\documentstyle[prl,aps,multicol]{revtex}
\begin{document}
\author{K.L. Sebastian and Alok K.R. Paul}
\address{Department of Inorganic and Physical Chemistry\\
Indian Instittue of Science\\
Bangalore 560012\\
India}
\title{BARRIER CROSSING BY A LONG CHAIN MOLECULE - \\
THE KINK MECHANISM}
\maketitle

\begin{abstract}
\large
We consider the generalization of the Kramers escape over a barrier problem
to the case of a long chain molecule. The problem involves the motion of a
chain molecule of $N$ segments across a region where the free energy per
segment is higher, so that it has to cross a barrier. We consider the limit
where the length of the molecule is much larger than the width of the
barrier. The width is taken to be sufficiently wide that a coninuum
description is applicable to even the portion over the barrier. We use the
Rouse model and analyse the mechanism of crossing a barrier. There can be
two dominant mechanisms. They are: end crossing and hairpin crossing. We
find the free energy of activation for the hairpin crossing is two times
that for end crossing. In both cases, the activation energy has a square
root dependence on the temperature $T$, leading to a non-Arrhenius form for
the rate. We also show that there is a special time dependent solution of
the model, which corresponds to a kink in the chain, confined to the region
of the barrier. The movement of the polymer from one side to the other is
equivalent to the motion of the kink on the chain in the reverse direction.
If there is no free energy difference between the two sides of the barrier,
then the kink moves by diffusion and the time of crossing $t_{cross}\sim
N^2/T^{3/2}$. If there is a free energy difference, then the kink moves with
a non-zero velocity from the lower free energy side to the other, leading to 
$t_{cross}\sim N/\sqrt{T}$. We also consider the translocation of
hydrophilic polypeptides across hydrophobic pores, a process that is quite
common in biological systems. Biological systems accompolish this by having
a hydrophobic signal sequence at the end that goes in first. We find that
for such a molecule, the transition state resembles a hook, and this is in
agreement with presently accepted view in cell biology.
\end{abstract}

\large

\section{Introduction}

The escape of a particle over a one dimensional barrier has been the subject
of a large number of investigations. The problem, often referred to as the
Kramers problem \cite{Kramers,Chandra}, has been the subject of detailed
reviews \cite{Hanggi,Melnikov}. Kramers found solutions in the limit of weak
friction and also in the limit of moderate to strong damping \cite{Chandra}.
The intermediate regime has been an active area of investigation \cite
{Hanggi}. The reason for this extensive activity is that this forms a model
for a chemical reaction occuring in a condensed medium. Kramers problem for
few degrees of freedom has also been the topic of study \cite
{Hanggi,Melnikov}. The quantum problem of escape/tunneling through a barrier
too is of considerable interest. In the case where the system has an
infinite number of degrees of freedom, this has been referred to as the
decay of metastable vaccum, a problem that has attracted quite a bit of
atttention in field theory, cosmology and mesoscopic quantum phenomena \cite
{Coleman,Kleinert}. In this paper, we consider a similar situation involving
only classical physics. The trapped object has $N$ $\left( \rightarrow
\infty \right) $ degrees of freedom, and is a polymer (a string). Though
there are no quantum effects, the problem is similar, and some experiments
are already available, that the results of the theory are easily verified.
Further, the mathematics is considerably simpler than in the other cases,
being equivalent to that of quantum mechanical tunneling of a single
particle in a bistable potential.

The way that the $N$ degrees of freedom are connected (a chain or a string)
leads to interesting new aspects to the problem that are not present in the
case where there are only finite number of degrees of freedom. Also, the
problem is of great interest in biology as many biological processes involve
the translocation of a chain molecule from one side of a membrane to the
other, through a pore in the membrane. The translocation of proteins from
the cytosol into the endoplasmic reticulum, or into mitochondria or
chloroplasts are processes of great interest and importance. Often, the
proteins are hydrophilic and the pore in the membrane forms a hydrophobic
region, through which it has to pass through \cite{HR,parkprl,Alberts,Simon}%
, resulting in an increase in the free energy for the portion of the chain
inside the pore. In infection by bacteriophages, conjugative DNA transfer
etc, long chain DNA molecules snake through pores in membranes \cite
{Dreise,Citovsky}. In all these cases, the chain molecule seems to get
across the membrane rather easily, contrary to the expectation that one gets
from the theoretical analysis available in the literature on the subject
(see below). Bezerukov et. al. \cite{Bezerukov} have studied the
partitioning of polymer molecule into a nanoscale pore. Chipot and Pohoille%
\cite{CP} have carried out a molecular dynamics simulation of a polypeptide,
translocating through the interface between hexane and water. They found
that the polypeptide (undecamer of poly-L-leucine), initially placed in a
random coil conformation on the aqueous side of the interface rapidly
translocates to the interfacial region and then folds. In another
interesting experiment, Han et. al. \cite{Han} observed the forced movement
of long, double stranded DNA molecules through microfabricated channels
which have regions that present an entropic barrier for the entry of the
molecules.

All these problems involve the passage of a long chain molecule, through a
region in space, where the free energy per segment is higher, thus
effectively presenting a barrier for the motion of the molecule. This
problem forms the generalization that we refer to as the Kramers problem for
a chain molecule. On the theoretical side, a variety of studies exist on
this kind of problem. Muthukumar and Baumgartner \cite{MB} studied the
movement of self avoiding polymer molecules between periodic cubic cavities
seperated by bottlenecks. The passage through the bottleneck presents an
entropic barrier to the motion, and they show that it leads to an exponetial
slowing down of diffusion with the number of segments $N$ in the chain.
Baumgartner and Skolnick \cite{BS} studied the movement of polymers through
a membrane driven by an external bias and membrane asymmetry. Park and Sung 
\cite{parkprl,PS}, have studied the translocation through a pore. They
analyze the passage through a pore on a flat membrane, with only the effects
of entropy included. The resultant entropic barrier is rather broad, its
width being proportional to $N$. Consequently, they consider the
translocation process as being equivalent to the motion of the center of
mass of the molecule. Using the result of the Rouse model that the diffusion
coefficient of the center of mass is proportional to $1/N$, they effectively
reduce the problem to the barrier crossing of single particle having a
diffusion coefficient proportional to $1/N$. As the translocation involve
motion of $N$ segments across the pore, the time taken to cross, $t_{cross}$
scales as $N^3$. They also show that in cases where there is adsorption on
the trans side, translocation is favored and then $t_{cross}$ scales as $N^2$%
. In a very recent paper, Park and Sung \cite{parkkramers} have given a
detailed investigation of the dynamics of a polymer surmounting a potential
barrier. They use multidimensional barrier crossing theory to study the
motion of a chain molecule over a barrier, in the limit where the width of
the barrier is much larger than the lateral dimension of the molecule. In an
interesting recent paper, Lubensky and Nelson \cite{Lubensky} study a case
where they assume the interaction of the segments of the polymer with the
pore to be strong. They argue that effectively, the dynamics of the portion
of the chain inside the pore is the one that is important and this, they
show, can give rise to $t_{cross}$ proportional to $N$. Again, they assume
diffusive dynamics. In a recent paper, we have suggested \cite{Sebastian} a
kink mechanism for the motion of the chain across a barrier and it is our
aim to give details of this mechanism in this paper.

We consider a polymer undergoing activated crossing over a barrier. This can
form a model for a polymer going through a pore too, as the pore can cause
an increase in the free energy of the segments inside it, as they would
interact with the walls of the pore. The width $w$ of the barrier is assumed
to be much larger than the Kuhn length $l$ of the polymer, but small in
comparison with the total length $Nl$ of the polymer. That is, $l<<w<<Nl$ .
For example, in eukaryotic cells\cite{Alberts}, the length of the nuclear
pore is about 100 \AA , while the Kuhn length for a single stranded DNA is
perhaps around 15 \AA\ \cite{Lubensky}. Therefore, one is justified in using
a continuum approach to the dynamics of the long chain. (It is possible to
retain the discrete approach, and develop the ideas based on them, but this
is more involved mathematically). Our approach is the following: We describe
the motion of the polymer using the Rouse model. The force that the barrier
exerts on the chain appears as an additional, non-linear term in the model.
We refer to this as the non-linear Rouse model. The non-linear term causes a
distortion of the portion of the chain inside the barrier, which we refer to
as the kink. Movement of the chain across the barrier is equivalent to the
motion of the kink in the reverse direction. The kink is actually a special
solution of the non-linear Rouse model, arising because of the
non-linearity. In the presence of a free energy difference between the two
sides, the kink moves with a definite velocity and hence the polymer would
cross the barrier with $t_{cross}$ proportional to $N$. Traditionally, the
non-linear models that one studies (for example, the $\phi ^4$or the
sine-Gordon model \cite{solitons,RR,Scott,Chaikin}) have potentials that are
translationally invariant, and hence the kink can migrate freely in space.
In comparison, in our problem, the non-linear term is fixed in position
space and hence the kink too is fixed in space. However, the chain molecule
(modelled as a string \cite{DE}) can move in space and hence the kink
migrates, not in space, but on the chain. As far as we know, such a
suggestion has never been made in the past and we believe that this is a
very useful idea in understanding polymer translocation.

In general, the polymer can escape by essentially two mechanisms. The first,
which we refer to as end crossing, involves the passage of one end of the
polymer over the barrier, by thermal activation. This leads to the formation
of the kink, which is then driven the free energy difference between the two
sides of the barrier. The second is by the escape of any portion of the
polymer over the barrier, in the form of a hairpin. The hairpin is a
kink-antikink pair. For a flexible polymer, the hairpin crossing has twice
the activation energy for end escape and hence one expects it to be less
probable. However, as it can take place anywhere on the chain, the frequency
factor for it is proportional to $N$ and hence for a sufficiently long
chain, this can become the dominant mechanism for the escape. Hairpin
crossing leads to the formation of a kink-antikink pair. The pair moves
apart on the chain, driven by the free energy gain and hence the time of
crossing is still proportional to $N$, though one expects that it is roughly
half the time of crossing in the end crossing case. In principle, in
addition to these, it is possible for more than one hairpin to be formed.
However, it is obvious that in passage through a pore, unless the pore is
rather wide, only the end-crossing mechanism would operate.

In all our mathematical development, we use the one dimensional version of
the Rouse model. This is no limitation, if one is concerned with
translocation across the interface between two immiscible liquids or the
experiments of Han et. al.\cite{Han}, which involve motion in a channel,
whose width is large in comparison with the size of the molecule. On the
other hand, if one is interested in translocation through a pore, strictly
speaking, one has to consider the full three dimensional nature of the
problem, which at present seems rather involved. However, we believe that
the one dimensional model captures the essential physics of the problem. Our
analysis should also be useful in situtations where the whole of the polymer
is in a pore, so that the dynamics may be taken to be one dimensional, with
the chain trying to cross a region of high free energy.

\section{The Model}

\subsection{The Free Energy Landscape}

The considerations in this section are quite general and do not depend on
the model that one uses to describe the polymer dynamics but involves the
assumption that the polymer is flexible over a length scale comparable to
the width of the barrier. We start by considering the free energy landscape
for the crossing of the barrier. The barrier and the polymer stretched
across it are shown in the figure \ref{fig1}. The polymer has initially all
its units on the cis side, where its free energy per segment is taken to be
zero. So the initial state, corresponds to a free energy of zero in the free
energy hypersurface shown in figure \ref{fig2}. In crossing over to the
trans side, it has to go over a barrier, as in the figure \ref{fig1}. The
transition state for the crossing can be easily found, from physical
considerations. It is the state shown in figure \ref{fig3}. In it, the
configuration of the polymer is such that the end of the polymer on the
trans side is located exactly at the point on the trans side at which its
free energy per segment is zero, with the other end on the cis side, and the
chain is such that the free energy of the whole chain is a minimum. This is
so because if one moves the end either in the forward or in the backward
direction (and the rest of the chain adjusted so that the free energy of the
chain as a whole minimum), then the total free energy of the system would
decrease. Hence in the free energy hypersurface figure \ref{fig2}, the
configuration shown in the figure \ref{fig3} corresponds to the maximum
(i.e. transition state). Once the system has crossed the transition state,
the chain is stretched across the barrier. The path of steepest descent then
corresponds to moving segments from the cis side to the trans side, with out
changing the configuration of the polymer in the barrier region. As there is
a free energy difference $\Delta V$ between the two sides, this would lead
to a lowering of the free energy by $\Delta V$ per segment, and this leads
to a path on the free energy surface with a constant slope, and of width $W$
proportional to $N$ (see figure 3). Such a landscape implies that the
translocation process would involve two steps. First step is going through
the transition state by the overcoming of the activation barrier. Once the
system has done this, it encounters a rather wide region of width
proportional to the length of the chain. Crossing this region is the second
step. As this region has a constant slope, the motion in this region is
driven and it is similar to that of a Brownian particle subject to a
constant force. Such a particle would take a time $t_{cross}$, proportional
to $N$ to cross this region.

Till now, we considered the case of end-crossing. The scenario for hairpin
crossing is similar. However, the activation energy is higher for hairpin
crossing. In hairpin crossing, the transition state is equivalent to the one
end crossing, repeated two times. Hence the activation energy for the
process is two times larger. Once a hairpin crossing occur, a kink-antikink
pair is formed and the kink and the anti-kink separate rapidly, due to the
driving force of the free energy gain. Then futher crossing occurs by the
movement of these two on the chain, which again leads to a time of crossing
proportional to $N$.

In the following we make all these considerations quantitative, using the
Rouse model to describe the dynamics of the chain.

\subsection{The Dynamics}

We consider the continuum limit of the Rouse model, discussed in detail by
Doi and Edwards \cite{DE}. The chain is approximated as a string, with
segments (beads) labelled by their position $n$ along the chain. $n$ is
taken to be a continuous variable, having values ranging from $0$ to $N$.
The position of the $n^{th}$ segment in space is denoted by $R(n,t)$, where $%
t$ is time. In the Rouse model, this position undergoes overdamped Brownian
motion and its time development is described by the equation

\begin{equation}
\label{one}\zeta \frac{\partial R(n,t)}{\partial t}=m\frac{\partial ^2R(n,t)%
}{\partial n^2}-V^{\prime }(R(n,t))+f(n,t). 
\end{equation}
In the above, $\zeta $ is a friction coefficient for the $n^{th}$ segment.
The term $m\frac{\partial ^2R(n,t)}{\partial n^2}$ comes from the fact that
stretching the chain can lower its entropy and hence increase its free
energy. Consequently, the parameter $m=3k_BT/l^2$ (see Doi and Edwards \cite
{DE}, equation (4.5). They use the symbol $k$ for the quantity that we call $%
m$) . As the ends of the string are free, the boundary conditions to be
satisfied are $\left\{ \frac{\partial R(n,t)}{\partial n}\right\}
_{n=0}=\left\{ \frac{\partial R(n,t)}{\partial n}\right\} _{n=N}=0$. $V(R)$
is the free energy of a segment of chain, located at the position $R$. We
assume that $V(R)$ leads to a barrier located near $R=0$. $f(n,t)$ are
random forces acting on the $n^{th}$ segment and have the correlation
function $\left\langle f(n,t)f(n_1,t_1)\right\rangle =2\zeta k_BT\delta
(n-n_1)\delta (t-t_1)$(see \cite{DE}, equation (4.12)). The deterministic
part of the equation (\ref{one}), which will play a key role in our
analysis, is obtained by neglecting the random noise term in (\ref{one}). It
is:

\begin{equation}
\label{two}\zeta \frac{\partial R(n,t)}{\partial t}=m\frac{\partial ^2R(n,t)%
}{\partial n^2}-V^{\prime }(R(n,t)) 
\end{equation}
This may also be written as:

\begin{equation}
\label{three}\zeta \frac{\partial R(n,t)}{\partial t}=-\frac{\delta E[R(n,t)]%
}{\delta R(n,t)} 
\end{equation}
where $E[R(n,t)]$ is the free energy functional for the chain given by:

\begin{equation}
\label{four}E[R(n,t)]=\int_0^Ndn\left[ \frac m2\left( \frac{\partial R(n,t)}{%
\partial n}\right) ^2+V(R(n,t))\right] 
\end{equation}

\subsection{The form of the barrier}

The chain is assumed to be subject to a biased double well potential (BDW),
of the form shown in the figure 1. The two minima are at $-a_0$ and $a_1$,
with $a_0<a_1$. There is assumed to be a maximum at $R=0$. Further, we take $%
V(-a_0)=0$. All these conditions can be satisfied if one takes $V^{\prime
}(R)=2\,k\,R\left( R+a_0\right) \,\left( R-a_1\right) $. Here, $k$ is a
constant and will determine the height of the barrier. Integrating this and
using $V(-a_0)=0$, we get 
\begin{equation}
\label{five}V(R)=\frac k6(R+a_0)^2(3R^2-2Ra_0-4Ra_1+a_0^2+2a_0a_1) 
\end{equation}
The barrier height for the forward crossing is $V_f=V(0)-V(-a_0)=\frac{\,1}6%
ka_0{}^3\,\left( a_0+2\,a_1\right) $ and for the reverse process, it is $%
V_b=V(0)-V(a_1)=\frac{1\,}6ka_1{}^3\left( 2\,a_0+a_1\right) $. On crossing
the barrier, a unit of the polymer lowers its free energy by $\Delta
V=V(a_1)-V(-a_0)=\frac{1\,}6k\left( a_0-a_1\right) \,\left( a_0+a_1\right)
^3 $. The form of the potential is shown in the figure \ref{fig3}.

\subsection{The Activation Free Energy for End and Hairpin Crossings \label
{activation}}

In this section, we consider the first step and calculate the activation
free energy for both end and hairpin crossing. Activation free energy can be
obtained from the free energy functional of equation (\ref{two}). This free
energy functional implies that at equilibrium, the probability distribution
functional is $\exp \left[ -\frac 1{k_BT}\int dn\left\{ \frac 12m\left( 
\frac{dR}{dn}\right) ^2+V(R(n))\right\} \right] $. The configurations of the
polymer which makes free energy a minimum are found from $\frac{\delta
E[R(n)]}{\delta R(n)}=0$, which leads to the equation

\begin{equation}
\label{six}m\frac{d^2R}{dn^2}=V^{\prime }(R) 
\end{equation}
Notice that this is just a Newton's equation for a particle (ficticious,
ofcourse) of mass $m$ moving in a potential $-V(R)$. This equation has four
solutions that are of interest to us. The first two are: (1) $R(n)=-a_0$,
(2) $R(n)=a_1$ which are the minima of the free energy. The first solution
is the initial state, where the polymer is trapped in the vicinity of $-a_0$%
. The second is the most stable minimum, at $R(n)=a_1$. In addition to
these, there are two more solutions which are of interest to us. These are $%
n $ dependent and correspond to end and hairpin crossings.

\subsubsection{End Crossing}

As we are interested in the case where the polymer is very long, we can
imagine $n$ to vary from $-\infty $ to $0$ and find a saddle point in the
free energy surface by searching for a solution satisfying $R(-\infty )=-a_0$
and the other end of the polymer to be at a point with $R>R_{\max }$, where $%
R_{\max }$ is the point where $V(R)$ has its maximum value. For the Newton's
equation (\ref{six}) the conserved energy is $E_c=\frac 12m\left( \frac{dR}{%
dn}\right) ^2-V(R(n))$. For the extremum path, $E_c=0$. Thus, the particle
starts at $R(-\infty )=-a_0$ with the velocity zero (this follows from the
boundary conditons of the Rouse model) and ends up at $R_f$ at the ''time'' $%
n=0$. Here $R_f(>R_{\max }),$ is the point such that $V(R_f)=0$, again with
the velocity zero. Further, free energy of this configuration is activation
free energy for end crossing. As for this configuration, $\frac 12m\left( 
\frac{dR}{dn}\right) ^2=V(R(n))$, we find the activation free energy to be
given by

\begin{equation}
\label{seven}E_{a,end}=\int_{-a_0}^{R_f}\sqrt{2mV(R)}dR. 
\end{equation}
The end crossing is illustrated in figure \ref{fig3a}.

\subsubsection{Hairpin Crossing}

If one imagines $n$ to vary in the range $(-\infty ,\infty )$ a second
saddle point may be found by taking $R(-\infty )=-a_0$ and $R(\infty )=-a_0$%
, so that the Newtonian particle starts at $-a_0$, makes a round trip in the
inverted potential $-V(R)$ and gets back to its starting point. This
obviously has an activation energy

\begin{equation}
\label{eight}E_{a,hp}=2\int_{-a_0}^{R_f}\sqrt{2mV(R)}dR=2E_{a,end} 
\end{equation}
Thus the activation energy is exactly two times for end crossing\cite{text1}%
. The hairpin crossing is shown in figure \ref{fig3b}

\subsubsection{The Temperature dependence}

As the parameter $m$ is proportional to the temperature ( $=3k_BT/l^2$ ), we
arrive at the general conclusion that both the activation energies $%
E_{a,end} $ and $E_{a,hp}$ are proportional to $\sqrt{T}$. For our model
potential of equation (\ref{five}) we find $R_f=a_0(\gamma -\,\sqrt{\gamma
^2-\gamma })$ where $\gamma =(1+2\frac{a_1}{a_0})\frac 13$ and 
\begin{equation}
\label{three3}E_{a,end}=\frac{\sqrt{mk}a_0^3}6\left[ (3\,\gamma ^2+1)\sqrt{%
1+3\gamma }-3\,\gamma \,(\gamma ^2-1)\ln \left( \sqrt{\gamma (\gamma -1)}%
/\left( 1+\gamma -\sqrt{1+3\gamma }\right) \right) \right] . 
\end{equation}
The Boltzmann factor $e^{-\frac{E_{act}}{k_BT}}$ for the crossing of one end
of the polymer over the barrier thus has the form $e^{-\text{constant}/\sqrt{%
T}}$. Further, we find that it is independent of $N$ for large $N$.

\section{The Rate of Crossing}

\subsection{Hairpin Crossing}

We now calculate the rate of crossing in the two cases. We first consider
the hairpin crossing, as this has connections with material available in the
literature \cite{Kleinert}. The methods that we use are quite well known in
the soliton literature \cite{solitons} and hence we give just enough details
to make the approach clear. The Rouse model in the equation (\ref{one})
leads to the functional Fokker Planck equation

\begin{equation}
\label{nine}\frac{\partial P}{\partial t}=\frac 1\zeta \int_0^Ndn\frac \delta
{\delta R(n)}\left[ k_BT\frac{\delta P}{\delta R(n)}+\frac{\delta E[R(n)]}{%
\delta R(n)}P\right] 
\end{equation}
for the probability distribution functional $P$. This equation implies that
the flux associated with the co-ordinate $R(n)$ is\cite{solitons}

\begin{equation}
\label{ten}j(R(n))=-\frac 1\zeta \left[ k_BT\frac{\delta P}{\delta R(n)}+%
\frac{\delta E[R(n)]}{\delta R(n)}P\right] 
\end{equation}

We now consider the initial, metastable state. As the rate of escape is
small, we can assume the probability distribution to be the equilibrium one,
which is

\begin{equation}
\label{twelve}P=\frac 1{Z_0}\exp \left\{ -E[R(n)]/k_BT\right\} 
\end{equation}
To determine $Z_0$ we use the condition $\int D[R(n)]$ $P=1$, where $\int
D[R(n)]$ stands for functional integration. It is convenient to introduce
the normal co-ordinates for small amplitude motion around the metastable
minimum and do the functional integration using them. For this, we expand $%
E[R(n)]$ around the metastable minimum, by putting $R(n)=-a_0+\delta R(n)$,
and expanding as a functional Taylor series in $\delta R(n)$ and keeping
terms up to second order in $\delta R(n)$. Then

\begin{equation}
\label{thirteen}E[R(n)]=\frac 12m\int_0^N\delta R(n)\left( -\frac{\partial ^2%
}{\partial n^2}+\omega _0^2\right) \delta R(n) 
\end{equation}
We have defined $\omega _0$ by putting $m\omega _0^2=\left[ \frac{\partial
^2V(R)}{\partial R^2}\right] _{R=-a_0}$. The normal (Rouse) modes are just
the eigenfunctions $\psi _k(n)$ of the operator $\widehat{H}^{ms}=\left( -%
\frac{\partial ^2}{\partial n^2}+\omega _0^2\right) $, having the eigenvalue 
$\varepsilon _k$ and satisfying the Rouse boundary conditions $\frac{%
\partial \psi _k(n)}{\partial n}=0$ at the two ends of the string. (The
superscript ''ms'' in $\widehat{H}^{ms}$s$\tan $ds for metastable). Now we
can expand $\delta R(n)$ as $\delta R(n)=\sum_kc_k\psi _k(n)$ so that the
expression for energy (\ref{thirteen}) becomes

\begin{equation}
\label{fourteen}E[R(n)]=\frac 12m\sum_k\varepsilon _kc_k^2 
\end{equation}
We now do the functional integration using the variables $c_k$. Then the
normalization condition $\int D[R(n)]$ $P=1$ becomes $\frac 1{Z_0}%
\prod\limits_k$ $\int dc_k\exp \left[ -\frac 12m\beta \varepsilon
_kc_k^2\right] =1$. This leads to $Z_0=\prod\limits_k\left( \frac{2\pi }{%
m\beta \varepsilon _k}\right) ^{1/2}$. Now we consider the vicinity of a
saddle point, where the probability distribution deviates from the
equilibrium one. We first consider the saddle point which corresponds to
hairpin crossing. The potential of the equation (\ref{five}) is rather
difficult to handle as we have not been able to obtain analytic solutions to
the Newton's equation (\ref{six}). In determining the crossing of the
barrier, the key role is played by the quantities $\omega _0$ and the height
of the barrier for crossing in the forward direction $V_f$. The quantities
that we calculate in this section have no dependence of the behavior of the
potential near the stable minimum. So, instead of using the quartic
potential of the equation (\ref{five}), we use the simpler cubic potential
of equation (\ref{fifteen}). This has no stable minimum (corresponding to
the final state), but that does not matter, because the quantities that we
calculate do not depend on its existence. Thus we use the potential:

\begin{equation}
\label{fifteen}V_c(R)=V_0\left( \frac{R+a_0}{R_0}\right) ^2\left( 1-\frac{%
R+a_0}{R_0}\right) 
\end{equation}
where we adjust $V_0$ and $R_0$ to reproduce the values for $\omega _0$ and
the barrier height $V_f$. Solving the equation (\ref{six}) for this
potential, in the limit of an infinitely long chain extending from $%
n=-\infty $ to $+\infty $, the saddle point that corresponds to hairpin
crossing is easily found to be given by the equation 
\begin{equation}
\label{sixteen}R_{hp}(n)=-a_0+R_0\left\{ \sec h\left( \sqrt{\frac{V_0}{2m}}%
n\right) \right\} ^2 
\end{equation}
In fact one has a continuous family of solutions of the form $R_{hp}(n-n_0)$%
, where $n_0\in \left( -\infty ,\infty \right) $ is arbitrary and determines
center of the kink-antikink pair. Now expanding the energy $E[R(n)]$ about
this saddle, by writing $R(n)=R_{hp}(n-n_0)+$ $\delta R(n)$ we get

\begin{equation}
\label{seventeen}E[R(n)]=E_{a,hp}+\frac 12m\int dn\delta R(n)\left[ -\frac{%
\partial ^2}{\partial n^2}+\omega _0^2\left\{ 1-3\sec h^2\left( \omega
_0(n-n_0)/2\right) \right\} \right] \delta R(n) 
\end{equation}
For the potential of equation (\ref{fifteen}) $E_{a,hp}=\left(
8R_0/15\right) \sqrt{2mV_0}$. The normal modes for fluctuations around the
saddle are determined by the eigenfunctions of the operator $\widehat{H}%
^{\ddagger }=-\frac{\partial ^2}{\partial n^2}+\omega _0^2\left\{ 1-3%
%TCIMACRO{\limfunc{sech}}
%BeginExpansion
\mathop{\rm sech}
%EndExpansion
^2\left( \omega _0(n-n_0)/2\right) \right\} .$ ($\ddagger $ is used to
denote the saddle point). The eigenfunctions are: (a) the discrete states $%
\psi _0^{\ddagger }$, $\psi _1^{\ddagger }$ and $\psi _2^{\ddagger }$ having
the eigenvalues $\varepsilon _0^{\ddagger }=-5\omega _0^2/4$, $\varepsilon
_1^{\ddagger }=0$ and $\varepsilon _2^{\ddagger }=3\omega _0^2/4$ and (b)
the continuum of eigenstates with eigenvalues of the form $\varepsilon
_k^{\ddagger }=\omega _0^2+k^2$(more details are given in appendix A). We
denote the eigenfunctions by $\psi _k^{\ddagger }$. The existence of the
eigenvalue $\varepsilon _1^{\ddagger }=0$ comes from the freedom of the
kink-antikink pair to have its center anywhere on the chain (the hairpin can
be formed anywhere). In the following, $\sum\limits_k$ would stand for
summation over all the eigenstates, including both the discrete and
continuum states while a symbol like $\sum\limits_{k\neq 1}$ means that the
bound state $\psi _1^{\ddagger }$ is to be excluded from the sum. Now
writing $\delta R(n)=\sum\limits_{k\neq 1}c_k^{\ddagger }\psi _k^{\ddagger }$%
, we get

$$
E[R(n)]=E_{a,hp}+\frac 12m\sum\limits_{k\neq 1}\varepsilon _k^{\ddagger
}\left( c_k^{\ddagger }\right) ^2 
$$
We write the probability density near the saddle as

\begin{equation}
\label{eighteen}P=\frac{\theta (c_0^{\ddagger },c_1^{\ddagger }...)}{Z_0}%
\exp \left\{ -\frac{E[R(n)]}{k_BT}\right\} 
\end{equation}
where $\theta (c_0^{\ddagger },c_1^{\ddagger }...)$, is a function that must
approach unity in the vicinity of the metastable minimum. Near the saddle,
one can calculate the flux $j_k^{\ddagger }$ in the direction of $%
c_k^{\ddagger }$.

$$
j_k^{\ddagger }=-\frac 1\zeta \left[ k_BT\frac{\partial P}{\partial
c_k^{\ddagger }}+\frac{\partial E[R(n)]}{\partial c_k^{\ddagger }}P\right] 
$$
Using the equations (\ref{seventeen} ) and (\ref{eighteen} ) we get

\begin{equation}
\label{ninteen}j_k^{\ddagger }=-\frac{k_BT}{Z_0\zeta }\frac{\partial \theta
(c_0^{\ddagger },c_1^{\ddagger }...)}{\partial c_k^{\ddagger }}\exp \left\{ -%
\frac 1{k_BT}\left( E_{a,hp}+\frac 12m\sum\limits_{k\neq 1}\varepsilon
_k^{\ddagger }\left( c_k^{\ddagger }\right) ^2\right) \right\} 
\end{equation}
In a steady state, there is flux only in the unstable direction. That is,
only $j_0^{\ddagger }$ is non-zero. This means that $\theta $ can depend
only on $c_0^{\ddagger }$, which impies that $j_0^{\ddagger }$ must have the
form

\begin{equation}
\label{twenty}j_0^{\ddagger }=A\exp \left\{ -\frac 1{k_BT}\left( \frac 12%
m\sum\limits_{k>1}\varepsilon _k^{\ddagger }\left( c_k^{\ddagger }\right)
^2\right) \right\} 
\end{equation}
where $A$ is a constant, to be determined. Using the equation (\ref{twenty})
in (\ref{ninteen}) we get $\frac{\partial \theta (c_0^{\ddagger })}{\partial
c_0^{\ddagger }}=-A\exp \left\{ -\frac m{2k_BT}\left| \varepsilon
_0^{\ddagger }\right| \left( c_0^{\ddagger }\right) ^2\right\} $. The fact
that $\theta (c_0^{\ddagger })$ must approach unity as $c_0^{\ddagger
}\rightarrow -\infty $, enables one to get $A=\left( \frac{m\left|
\varepsilon _0^{\ddagger }\right| }{2\pi k_BT}\right) ^{1/2}$ . Hence $%
\theta (c_0^{\ddagger })=\left( \frac{m\left| \varepsilon _0^{\ddagger
}\right| }{2\pi k_BT}\right) ^{1/2}\int_{c_0^{\ddagger }}^\infty dz\exp
\left\{ -\frac 1{2k_BT}m\left| \varepsilon _0^{\ddagger }\right| z^2\right\}
.$ Now the net flux crossing the barrier is found by integrating $%
j_0^{\ddagger }$ over all directions other than $c_0^{\ddagger }$ . The
integrals over all $c_k^{\ddagger }$, except $c_1^{\ddagger }$ is
straightforward. As $\varepsilon _1^{\ddagger }=0$, $\int dc_1^{\ddagger }$
needs special handling. The integral, as is well-known, is performed by
converting it to an integral over the kink-antikink position, $n_0$. That
is, $\int dc_1^{\ddagger }=\alpha \int dn_0,$ where $\alpha ^2=\int_{-\infty
}^\infty dn\left( \frac{\partial R_{hp}(n)}{\partial n}\right) ^2=\frac{%
E_{a,hp}}m$. Hence the rate becomes

\begin{equation}
\label{twentya}k_{hp}=\frac{k_BT}{Z_0\zeta }\left( \frac{m\left| \varepsilon
_0^{\ddagger }\right| }{2\pi k_BT}\right) ^{1/2}\prod\limits_{k>1}\left( 
\frac{2\pi k_BT}{m\left| \varepsilon _k^{\ddagger }\right| }\right)
^{1/2}\left( \frac{E_{a,hp}}m\right) ^{1/2}N\exp \left(
-E_{a,hp}/k_BT\right) 
\end{equation}
The notation $\prod\limits_{k>1}$ is used to indicate product over all
eigenvalues of $\widehat{H}^{\ddagger }$, except the first two. On using the
expression for $Z_0$,

\begin{equation}
\label{twentyone}k_{hp}=\frac{k_BT}\zeta \left( \frac m{2\pi k_BT}\right)
^{3/2}I_{hp}\left( \frac{\left| \varepsilon _0^{\ddagger }\right| E_{a,hp}}{%
\left| \varepsilon _2^{\ddagger }\right| m}\right) ^{1/2}N\exp \left(
-E_{a,hp}/k_BT\right) 
\end{equation}
where $I_{hp}=\left( \frac{\prod\limits_k\varepsilon _k}{\prod\limits_{k>2}%
\varepsilon _k^{\ddagger }}\right) ^{1/2}$. This infinite product is
evaluated in the appendix B and is found to be $I_{hp}=\frac{15}2\omega _0^3$%
. This leads to

\begin{equation}
\label{twentytwo}k_{hp}=\frac{5Nm\omega _0^3}{4\pi \zeta }\left( \frac{%
15E_{a,hp}}{2\pi k_BT}\right) ^{1/2}\exp \left( -E_{a,hp}/k_BT\right) 
\end{equation}

\subsection{End Crossing}

In this case, the analysis is similar to the above. The operator $\widehat{H}%
^{\ddagger }$ is the same as earlier. However, there is an interesting
difference. In the hairpin case, the boundary conditions on $\psi
_k^{\ddagger }$ ( $\frac{d\psi _k^{\ddagger }}{dn}=0$, at the two ends) were
at $n=\pm \infty $, while in this case, they are at $n=0$ and at $n=\infty $
(i.e. the boundary value problem is now on the half-line). Due to this, one
has to rule out the odd $\psi _k^{\ddagger }$ that exists in the hairpin
case as they do not satisfy the Rouse boundary condition$\frac{d\psi
_k^{\ddagger }}{dn}=0$ at $n=0$. So we consider only the even solutions.
Thus the eigenvalue at zero is ruled out (which is quite alright as end
crossing can occur only at the end and not anywhere else, but we will put in
additional factor of 2 as it can occur at the two ends). The discrete
spectrum now has only the eigenvalues $\varepsilon _0^{\ddagger }=-5\omega
_0^2/4$, and $\varepsilon _2^{\ddagger }=3\omega _0^2/4$. The expression for
the rate is

\begin{equation}
\label{twentythree}k_{end}=\frac{k_BT}\zeta \left( \frac{m\left| \varepsilon
_0^{\ddagger }\right| }{2\pi k_BT}\right) ^{1/2}\widetilde{I}_{end}\exp
\left( -E_{a,end}/k_BT\right) 
\end{equation}
where $\widetilde{I}_{end}=\frac{\prod\limits_{k\neq 0}\left( \frac{2\pi k_BT%
}{m\varepsilon _k^{\ddagger }}\right) ^{1/2}}{\prod\limits_k\left( \frac{%
2\pi k_BT}{m\varepsilon _k}\right) ^{1/2}}$. In this product, there are $N-1$
terms in the numerator and $N$ terms in the denominator. One of the $N-1$
terms is the bound state with an eigenvalue $\varepsilon _2^{\ddagger
}=3\omega _0^2/4$. Separating this out from the product, one can write $%
\widetilde{I}_{end}=\left( \frac{2m}{3\pi k_BT\omega _0^2}\right)
^{1/2}I_{end}$, where $I_{end}=\left( \frac{\prod\limits_k\varepsilon _k}{%
\prod\limits_{k>2}\varepsilon _k^{\ddagger }}\right) ^{1/2}$. the evaluation
of this product involves some subtelity and is done in the Appendix B. The
result is

$$
k_{end}=\frac{5m\omega _0^2}{2\sqrt{2}\pi \zeta }\exp \left(
-E_{a,end}/k_BT\right) 
$$
Accounting for the existence of two ends leads to

\begin{equation}
\label{twentyfour}k_{two-ends}=\frac{5m\omega _0^2}{\sqrt{2}\pi \zeta }\exp
\left( -E_{a,end}/k_BT\right) 
\end{equation}

\section{The kink and its motion}

\subsection{The kink solution and its velocity}

Having overcome the activation barrier, how much time would the polymer take
to cross it? We denote this time by $t_{cross}$. To calculate this, we first
look at the mathematical solutions of the deterministic equation (\ref{two}%
). The simplest solutions of this equation are: $R(n,t)=-a_0$ and $%
R(n,t)=a_1 $. These correspond to the polymer being on either side of the
barrier and these are just mean values of the position on the two sides.
Thermal noise makes $R(n,t)$ fluctuate about the mean position which may be
analyzed using the normal co-ordinates for fluctuations about this mean
position. Each normal mode obeys a Langevin equation similar to that for a
harmonic oscillator, executing Brownian motion. In addition to these two
time independent solutions, the above equation has a time dependent solution
(a kink) too, which corresponds to the polymer crossing the barrier. We
analyze the dynamics of the chain, with the kink in it, using the normal
modes for fluctuations about this kink configuration. Our analysis makes use
of the techniques that have been used to study the diffusion of solitons\cite
{solitons}

As is usual in the theory of non-linear wave equations, a kink solution
moving with a velocity $v$ may be found using the ansatz $R(n,t)=R_s(\tau )$
where $\tau =n-vt$ \cite{solitons}. Then the equation (\ref{two}) reduces to

\begin{equation}
\label{twentyfive}m\,\frac{d^2R_s}{d\tau ^2}+v\,\zeta \,\frac{dR_s}{d\tau }%
=V^{\prime }(R_s). 
\end{equation}

If one imagines $\tau $ as time, then this too is a simple Newtonian
equation for the motion of particle of mass $m$, moving in the upside down
potential $-V(R)$. However, in this case, there is a frictional term too,
and $v\,\zeta /m$ is the coefficient of friction. This term makes it
possible for us to find a solution for quite general forms of potential,
with $V^{\prime }(R)\rightarrow 0$ as $R\rightarrow \pm \infty $. For the
potential of the equation (\ref{six}), we can easily find a solution of this
equation, obeying the conditions $R_s(\tau )=-a_0$ for $\tau \longrightarrow
-\infty $ and $R_s(\tau )=a_1$ for $\tau \longrightarrow \infty .$ The
solution is

\begin{equation}
\label{twentysix}R_s(\tau )=\left( -a_0+e^{\tau \,\omega \,\left(
a_0+a_1\right) }\,a_1\right) \left( 1+e^{\tau \omega \,\left( a_0+a_1\right)
}\right) ^{-1}, 
\end{equation}
with $\omega \,=\sqrt{k/m}$ . The solution exists only if the velocity{\ }$v=%
\frac{\sqrt{mk}\,}\zeta \,(a_0-a_1).$ This solution is a kink, occurring in
the portion of the chain inside the barrier. We shall refer to the point
with $\tau =0$ as the center of the kink. (Actually one has a one-parameter
family of solutions of the form $R_s(\tau +\tau _0)$, where $\tau _0$ is any
arbitrary contant). As $\tau =n-vt$, the center of the kink moves with a
constant velocity $v$. Note that this velocity depends on the shape of the
barrier. Thus for our model potential, if $a_0<a_1$, then $V_f<V_b$, and
this velocity is negative. This implies that the kink is moving in the
negative direction, which corresponds to the chain moving in the positive
direction. That is, the chain moves to the lower free energy region, with
this velocity. If the barrier is symmetric, then $a_0=a_1$( $V_f=V_b$) the
velocity of the kink is zero.

\subsection{Fluctuations about the kink}

We now analyze the effect of the noise term present in the equation (\ref
{one}). The center of the kink can be anywhere on the chain - which means
that the kink is free to move on the chain. Actually, as the position of the
kink is fixed in space, this means that the polymer is moving across the
barrier. The kink would also execute Brownian motion, due to the noise term.
The motion of the kink caused by the noise terms is a well studied problem
in the literature \cite{solitons} and one can make use of these methods.
Following `Instanton methods' of field theory \cite{RR}, we write 
\begin{equation}
\label{twentyseven}R(n,t)=R_s(n-a(t))+\sum_{p=1}^\infty X_p(t)\phi
_p(n-a(t),t) 
\end{equation}
We have allowed for the motion of the kink by taking the kink center to be
at $a(t)$, where $a(t)$ is a random function of time which is to be
determined. $\phi _p$ are a set of functions (the Rouse modes) below and $%
X_p(t)$ are expansion coefficients. This may be put into the equation (\ref
{one}) to derive an equation of motion for $a(t).$ Neglecting kink-phonon
scattering leads to \cite{Sebastian}

\begin{equation}
\label{twentyeight}\stackrel{\cdot }{a}(t)=v+\xi _0(t)/C 
\end{equation}
where we define $\psi _0(n)$ by $\partial _nR_s(n)=C\psi _0^{*}(n)$ with 
\begin{equation}
\label{twentynine}C^2=\left\langle \partial _nR_s(n)\left| e^{v\zeta 
\overline{n}/m}\right| \partial _nR_s(n)\right\rangle =\frac{2\,}3\pi
\,\omega \,\csc (2\,\pi \frac{\,a_1-a_0}{a_0+a_1})\,\left( a_1-a_0\right)
\,a_0\,a_1. 
\end{equation}
and

\begin{equation}
\label{thirty}\xi _0(t)=\frac 1\zeta \int_{-N/2}^{N/2}dn\psi
_0^{*}(n)e^{v\zeta \overline{n}/(2m)}f(n+a(t),t). 
\end{equation}
$\xi _0(t)$ is a random function of time, having the correlation function

\begin{equation}
\label{thirtyone}\left\langle \xi _0(t)\xi _0(t_1)\right\rangle =\delta
(t-t_1)(2k_BT/\zeta )\int_{over\text{ }the\text{ }kink}dne^{vn\zeta
/m}\left[ \psi _0(n)\right] ^2 
\end{equation}
.For the potential given by the equation (\ref{five}) one gets

\begin{equation}
\label{thirtytwo}\left\langle \xi _0(t)\xi _0(t_1)\right\rangle =\delta
(t-t_1)k_BT/(2\zeta a_0\,a_1)\sec (2\,\pi \frac{\,a_1-a_0}{a_0+a_1}%
)\,\,\left( 3\,a_1-a_0\right) \,\left( 3\,a_0-a_1\right) . 
\end{equation}

The equations (\ref{twentyeight}) and (\ref{twentynine}) imply that the kink
position $a(t)$ executes Brownian motion with drift. As $v$ is negative, the
drift is in the negative direction.

\subsection{The crossing time $t_{cross}$}

For the polymer to cross the barrier, the kink has to go in the reverse
direction, by a distance equal to $N$. As the equation (\ref{twentyeight})
is just that for a particle executing Brownian motion with drift, we can
estimate the time of crossing as a first passage time. As the kink starts at
one end, we take the initial position of the particle, $a$ to be $N$ and
calculate the average time required for it to attain the value $0$, which
would correspond to the polymer crossing the barrier fully. Writing the
diffusion equation for the survival probability $P(a,t)$ for a particle
starting at $a=N$ at the time $t=0$ and being absorbed at $a=0$, we get

\begin{equation}
\label{thirtythree}\frac{\partial P(a,t)}{\partial t}=D\frac{\partial
^2P(a,t)}{\partial a^2}-v\frac{\partial P(a,t)}{\partial a} 
\end{equation}
. Here, the diffusion coefficient

$$
D=\frac 1{2tC^2}\int_0^tdt_1\int_0^tdt_2\left\langle \xi _0(t_1)\xi
_0(t_2)\right\rangle 
$$

\begin{equation}
\label{thirtyfour}=\frac{3\,k_BT\,\,\,}{8\,\pi \,\zeta \,}\sqrt{\frac mk}%
\frac{\left( 3\,a_1-a_0\right) \,\left( 3\,a_0-a_1\right) }{%
\,a_0{}^2\,a_1{}^2\,\left( a_1-a_0\right) }\tan (2\pi \frac{a_1-a_0}{a_0+a_1}%
). 
\end{equation}
The equation (\ref{thirtythree}) is to be solved, subject to the initial
condition $P(a,0)=\delta (a-N)$ and with absorbing boundary condition at $%
a=0 $ (i.e. $P(0,t)=0$) and $P(\infty ,t)=0$. It is easy to solve the above
equation in the Laplace domain. The result for the Laplace transform $%
\overline{P}(a,s)=\int_0^\infty dtP(a,t)\exp (-st)$ is:

\begin{equation}
\label{thirtyfive}\overline{P}(a,s)=\frac 1{\sqrt{4\,D\,s+v^2}}\left[ e^{%
\frac{\left( a-N\right) \,v-\sqrt{4\,D\,s+v^2}\,\left| a-N\right| }{2\,D}%
}-e^{\frac{\left( a-N\right) \,v-\sqrt{4D\,s+v^2}\,\left| a\right| -\sqrt{%
4\,D\,s+v^2}\,N}{2\,D}}\right] 
\end{equation}
The Laplace transform of the survival probability is given by $\overline{P}%
(s)=\int_{-\infty }^\infty da\overline{P}(a,s)$ and is found to be

\begin{equation}
\label{thirtysix}\overline{P}(s)=\frac 1s\left[ 1-e^{\frac{-N\,v-\sqrt{%
4\,D\,s+v^2}\,N}{2\,D}}\right] 
\end{equation}
The average crossing time is given by $t_{cross}=Limit_{s->0}\overline{P}%
(s)=N/(-v)\,$, if $v<0$. As $v$ is proportional $\sqrt{mk}$, assuming $V(R)$
to be temperature independent we find $t_{cross}\sim N/\sqrt{T}$. This is a
general conclusion, independent of the model that we assume for the
potential. If the barrier is symmetric, the kink moves with an average
velocity $v=0$. Taking the $v\rightarrow 0$ limit of $\overline{P}(s)$, we
get

\begin{equation}
\label{thirtyseven}\overline{P}(s)=\frac 1s\left( 1-e^{-\frac{\sqrt{s}\,N}{%
\sqrt{D}}}\right) 
\end{equation}
so that the survival probability becomes

\begin{equation}
\label{thirtyeight}P(t)=%
%TCIMACRO{\limfunc{Erf}}
%BeginExpansion
\mathop{\rm Erf}
%EndExpansion
(\frac N{2\,\sqrt{Dt}\,}). 
\end{equation}
This expression for the survival probability implies that the average time
that the particle survives is $t_{cross}\sim N^2/D.$ For the symmetric
barrier, the value of $D$ may be obtained by taking the limit $%
a_1\rightarrow a_0$, and one finds $D=\frac{3\,k_BT\,}{4\,\zeta \,\,a_0{}^3}%
\sqrt{\frac mk}$ and thus $t_{cross}\sim N^2/T^{3/2}$.

In their analysis, Park and Sung \cite{PS} considered the passage of a
polymer through a pore for which the barrier is entropic in origin.
Consequently it is very broad, the width being of the order of $N$. Hence
they consider the movement as effectively that of the center of mass of the
polymer which diffuses with a coefficient proportional to $1/N$. As the
center of mass has to cover a distance $N$, the time that it takes is
proportional to $N^3$. If there is a free energy difference driving the
chain from one side to the other, then the time is proportional to $N^2$. In
comparison, we take the barrier to be extrinsic in origin and assume its
width to be small in comparison with the length of the chain. The crossing
occurs by the motion of the kink, which is a localized non-linear object in
the chain whose width is of the same order as that of the barrier. As the
polymer is intially subject to a potential well, the entropic contribution
to the barrier that Park and Sung \cite{PS} consider does not exist in our
case. Such a potential is realistic, in cases where the polymer is subjected
to a driving force (for example an electric field). As the kink is a
localized object, its diffusion coefficient has no $N$ dependence and our
results are different from those of Park and Sung \cite{PS}. In the case
where there is no free energy difference, our crossing time is proportional
to $N^2$(in contrast to $N^3$ of Park and Sung) , while if there is a free
energy difference, our crossing time is proportional to $N$ (in contrast to $%
N^2$ of Park and Sung). In a very recent paper \cite{parkkramers}, Park and
Sung have considered the Rouse dynamics of a short polymer surmounting a
barrier. The size of the polymer is assumed to be small in comparison with
the width of the potential barrier. Consequently, the transition state has
almost all the units at the top of the barrier, leading to the prediction
that the activation energy is proportional to $N$. This leads to a crossing
probability that decreases exponentially with $N$. In comparison, as found
in section \ref{activation}, the free energy of activation does not depend
on the length of the chain. Hence, the mechanism is the favoured one for
long chains.

\subsection{The net rate}

As the actual crossing is a two step process, with activation as the first
step and kink motion as the second step, the net rate of the two has to be a
harmonic mean of the two rates. For a very long chain, the motion of the
kink has to become rate determining. In the case of translocation of
biological macromolecules, considered in section \ref{BIO} there does not
seem to be any free energy of activation and then the rate is determined by $%
t_{cross}$ alone. Recently, the motion of long chains in microfabricated
channels have been investigated by Han et al \cite{Han}. In contrast to the
situtation for a pore, there is an additional direction is available for the
molecule to form a hairpin, viz. perpendicular to the direction of movement
of the molecule. Consequently, in overcoming the barrier, both end crossing
and hairpin crossing can occur (see figures \ref{fig3a} and \ref{fig3b}).
Experimental results show that the longer molecule crosses the barrier
faster. This means that the $N$-dependence of $k_{hp}$ causes the hairpin
crossing to be the dominant mechanism of crossing in these experiments.

\section{How do biological systems lower the activation energy?\label{BIO}}

If there was a high activation energy ($>>k_BT$) for the translocation, the
process would be unlikely and hence, biological systems would not be able to
function, if they depended crucially on such transfers. As translocation
seem to be very efficient in biological systems, one needs to look at the
mechanism that evolution has designed to reduce the barrier. The destination
(referred to as sorting) of a biological long chain molecule is determined
by a sequence of units at the begining of the chain, referred to as the
signal sequence. For example, proteins destined to the endoplasmic reticulum
possess an amino-terminal signal sequence, while those destined to remain in
the cytosol do not have this. If one attaches this sequence to a cytosolic
protein, then the protein is found to end up in the endoplasmic reticulum
(see reference \cite{Alberts}, figure 14.6). The way the sequence works is
simple. If the pore is hydrophobic and the chain hydrophilic, then the
signal sequence is hydrophobic, so that the signal sequence has a low free
energy inside the pore.

We qualitatively analyze this type of problem in the following, using the
Rouse model. The way to model the situation would be to have a potential
that is dependent upon the segment number $n$ in the chain. Hence, in the
equations of the Rouse model the potential term would have an explicit
dependence on $n$. Let us denote the length of the signal sequence by $s$.
The simplest model would be to have a potential which is attractive, for $%
0<n<s$ and which has the shape of a barrier for $s<n<N$. The transition
state is determined by the the Newton-like equation

\begin{equation}
\label{thirtynine}m\frac{d^2R}{dn^2}=V_{new}^{\prime }(n,R),
\end{equation}
with $n$ playing the role of time (in the following we shall refer to $n$ as
the time for the motion of this ficticious particle). We take the potential
to be such that

$$
V_{new}(n,R)=-V(R)\text{ if }0<n<s\text{ and } 
$$

$$
=V(R)\text{ if }s<n<N\text{ } 
$$
This corresponds to a particle moving in a time dependent potential, which
switches from being repulsive to attractive at the time $s$. The shape of
this time dependent potential is shown in the figure \ref{fig6}. The
boundary conditions $\left\{ \frac{dR(n)}{dn}\right\} _{n=0}=\left\{ \frac{%
dR(n)}{dn}\right\} _{n=N}=0$ imply that the particle has to start and end
with zero velocity. Let us imagine that the particle starts at the point $%
R_0 $ (see figure \ref{fig6}). As the potential that it feels up to the time 
$s$ is repulsive, it follows the path indicated by the dashed line in the
figure, and the conservation of energy may be written as: $\frac 12m\left( 
\frac{dR(n)}{dn}\right) ^2+V(R)=V(R_0)$. Let it reach the point $R_s$ after
a time $s$. At this time, the potential is switched from $V(R)$ to $-V(R)$.
From this time on, the equation of conservation of energy would be:

\begin{equation}
\label{forty}\frac 12m\left( \frac{dR(n)}{dn}\right) ^2-V(R)=V(R_0)-2V(R_s).
\end{equation}

This is the equation of motion of the particle for $s<n<N$. We are
interested in $N\rightarrow \infty $ limit and we have to satisfy the
boundary condition $\left\{ \frac{dR(n)}{dn}\right\} _{n=N}=0$ at the end of
the chain. In the particle picture, this is equivalent to the condition that
the total energy of the particle obeying the equation \ref{thirtynine} must
be zero. This implies that $V(R_0)=2V(R_s)$. For a given $s$, this uniquely
fixes the values of the two variables $R_0$ and $R_s$.

The net transition state is shaped like a hook and the hydrophobic part of
the chain is completely in the short arm of the hook (see figure \ref{fig7})
. A configuration like the one in the figure \ref{fig8} where the whole of
the hook is formed by the hydrophobic part is not a transition state. The
transition state in figure \ref{fig7}, though it seems likely to occur in
crossing between liquid liquid interfaces, it seems rather difficult to form
in the case of passage through a pore as there are two difficulties: (1) the
chain has to bend to form the hook (2) the pore has to be wide enough to
accommodate the two strands of the hook simultaneously. Inspite of these,
nature does seem to use this as an inspection of the figure 14-14 of
reference \cite{Alberts} shows.

\section{Conclusions}

We have considered the generalization of the Kramers escape over a barrier
problem to the case of a long chain molecule. It involves the motion of
chain molecule of $N$ segments across a region where the free energy per
segment is higher, so that it has to cross a barrier. We consider the limit
where the width of the barrier $w$ is large in comparison with the Kuhn
length $l$, but small in comparison with the total length $Nl$ of the
molecule. The limit where $Nl<<w$ has been considered in a recent paper by
Park and Sung \cite{parkkramers}. We use the Rouse model and find there are
two possible mechanism that can be important - end crossing and hairpin
crossing. We calculate the free energy of activation for both and show that
both have a square root dependence on the temperature $T$, leading to a
non-Arrhenius form for the rate. We also find that the activation energy for
hairpin crossing is two times the activation energy for end-crossing.
Inspite of this, for long enough chains, where the geometry of the systems
permits, hairpin formation can be the dominant mode of escape as seen in the
experiments of Han et. al.\cite{Han}. The width of the barrier in these
experiments is rather large in comparison with the length of the polyme so
that the kink mechanism of crossing seems to be unlikely in this case.

While in the short chain limit Park and Sung find the activation energy to
be linearly dependent on $N$, we find that for long chains, the activation
energy is independent of $N$. We also show that there is a special time
dependent solution of the model, which corresponds to a kink in the chain,
confined to the region of the barrier. In usual non-linear problems with a
kink solution, the problem has translational invariance and the soliton/kink
can therefore migrate. In our problem, the translational invariance is not
there, due to the presence of the barrier and the kink solution is not free
to move in space. However, the polymer on which the kink exists, can move,
though the kink is fixed in space. Thus, the polymer goes from one side to
the other by the motion of the kink in the reverse direction on the chain.
If there is no free energy difference between the two sides of the barrier,
then the kink moves by diffusion and the time of crossing $t_{cross}\sim
N^2/T^{3/2}$. If there is a free energy difference, then the kink moves with
a non-zero velocity from the lower free energy side to the other, leading to 
$t_{cross}\sim N/\sqrt{T}$. We also consider the translocation of
hydrophilic polypeptides across hydrophobic pores. Biological systems
accompolish this by having a hydrophobic signal sequence at the end that
goes in first. Our analysis leads to the conclusion that for such a
molecule, the configuration of the molecule in the transition state is
similar to a hook, and this is in agreement with presently accepted view in
cell biology \cite{Alberts}. It is also possible that a kink movement
mechanism might operate in other biological phenomena, like protein folding%
\cite{Bose}

\section{Acknowledgements}

K.L. Sebastian is deeply indebted to Professor K. Kishore for the
encouragement that he has given over the years and this paper is dedicated
to his memory. He thanks Professors S. Vasudevan and Diptiman Sen for
discussions and Professors B. Cherayil and Indrani Bose for their comments.

\setcounter{section}{0} \setcounter{equation}{0} 
\renewcommand{\thesection}{\Alph{section}}

\section{appendix}

\renewcommand{\theequation}{\thesection -\arabic{equation}} 
\renewcommand{\thesubsection}{\Roman{subsection}}

\subsection{The eigenfunctions of the Hamiltonian $\widehat{H}^{\ddagger }$%
\label{eigen}}

The Hamiltonian $\widehat{H}^{\ddagger }=-\frac{\partial ^2}{\partial n^2}%
+\omega _0^2\left\{ 1-3%
%TCIMACRO{\limfunc{sech}}
%BeginExpansion
\mathop{\rm sech}
%EndExpansion
^2\left( \omega _0n/2\right) \right\} $ has the following eigenfunctions
(functions are not normalised) and eigenvalues, if $n$ allowed to be in the
range $\left( -\infty ,\infty \right) $.

\subsubsection{Discrete States}

1) $\psi _0(n)=%
%TCIMACRO{\limfunc{sech}}
%BeginExpansion
\mathop{\rm sech}
%EndExpansion
(\frac{\omega _0\,n}2)^3;$ \ $\varepsilon _0=-5\omega _0^2/4$

2) $\psi _1(n)=%
%TCIMACRO{\limfunc{sech}}
%BeginExpansion
\mathop{\rm sech}
%EndExpansion
(\frac{\omega _0\,n}2)^2\,\tanh (\frac{\omega _0\,n}2);$ \ $\varepsilon _1=0$

3) $\psi _2(n)=\left\{ -3+2\,\cosh (\omega _0\,n)\right\} \,%
%TCIMACRO{\limfunc{sech}}
%BeginExpansion
\mathop{\rm sech}
%EndExpansion
(\frac{\omega _0\,n}2)^3;$ \ $\varepsilon _2=3\omega _0^2/4$

\subsubsection{Continuum States}

The continous part of the spectrum starts at $\omega _0^2$. The potential is
reflectionless. Corresponding to an eigenvalue $\omega _0^2+k^2$, there are
two eigenfunctions, which we write as an odd function and an even function.
They are:

1) $\psi _{even}(n)=8\,k\,\left( k^2+\omega _0^2\right) \,\cos
(k\,n)-3\,\omega _0\,\left( 8\,k^2+3\,\omega _0^2\right) \,\sin
(k\,n)\,\tanh (\frac{\omega _0\,n}2)$

$-30\,k\,\omega _0^2\,\cos (k\,n)\,\tanh (\frac{\omega _0\,n}2)^2+15\,\omega
_0^3\,\sin (k\,n)\,\tanh (\frac{\omega _0\,n}2)^3$

2) $\psi _{odd}(n)=-8\,k\,\left( k^2+\omega _0^2\right) \,\sin
(k\,n)-3\,\omega _0\,\left( 8\,k^2+3\,\omega _0^2\right) \,\cos
(k\,n)\,\tanh (\frac{\omega _0\,n}2)$

$+30\,k\,\omega _0^2\,\sin (k\,n)\,\tanh (\frac{\omega _0\,n}2)^2+15\,\omega
_0^3\,\cos (k\,n)\,\tanh (\frac{\omega _0\,n}2)^3$

In the limit $n\rightarrow \pm \infty $, the even function becomes like

$\psi _{even}(n)=2\,k\,\left( 4\,k^2-11\,\omega _0^2\right) \,\cos (k\,x)\pm
6\,\omega _0\,\left( -4\,k^2+\omega _0^2\right) \,\sin (k\,x)$, which may be
written as $(Constant)\cos (kx\mp \delta (k))$, so that the phase shift $%
\delta (k)=\arctan (\frac{-3\omega _0\,\left( \omega _0^2-4\,k^2\right) }{%
k\,\left( -11\omega _0^2+4\,k^2\right) })$. The phase shift for the odd
solution is just the same. Hence the total change in the density of states
is given by $\Delta n(k)=\frac 2\pi \frac{d\delta (k)}{dk}=-\frac 2\pi
\left( \frac{\omega _0}{k^2+\omega _0^2}+\frac{2\,\omega _0}{4\,k^2+\omega
_0^2}+\frac{6\,\omega _0}{4\,k^2+9\,\omega _0^2}\right) $. On integration, $%
\int_0^\infty dk\Delta n(k)=-3$ as it should be, as there are three bound
states for $\widehat{H}^{\ddagger }$.

\subsection{Evaluation of the Infinite Products}

\subsubsection{Hairpin Crossing}

The infinite product is:

\begin{equation}
\label{A1}I_{hp}=\left( \frac{\prod\limits_k\varepsilon _k}{%
\prod\limits_{k>2}\varepsilon _k^{\ddagger }}\right) ^{1/2} 
\end{equation}
where $\varepsilon _k$ represent the eigenvalues of the continuum states of
the hamiltonian $\widehat{H}^{ms}=\left( -\frac{\partial ^2}{\partial n^2}%
+\omega _0^2\right) $and $\varepsilon _k^{\ddagger }$are the eigenvalues of $%
\widehat{H}^{\ddagger }$, satisfying the boundary conditions at $n=\pm
\infty $. The above product involves only the continuum eigenvalues of the
two Hamiltonians. Now,

$$
\ln I_{hp}=\frac 12\left( \sum_k\ln \varepsilon _k-\sum_{k>2}\ln \varepsilon
_k^{\ddagger }\right) 
$$

$$
=\frac 12\int_0^\infty dk\ln \left( \omega _0^2+k^2\right) \left(
n(k)-n_{hp}^{\ddagger }(k)\right) 
$$
where the $n(k)$ stands for the density of states in the continuum, for the
Hamiltonian $\widehat{H}^s$and $n_{hp}^{\ddagger }(k)$ for the Hamiltonian $%
\widehat{H}^{\ddagger }$. The change in the density of states is $\Delta
n_{hp}(k)=-n(k)+n_{hp}^{\ddagger }(k)$ and is easily evaluated from the
information given in subsection \ref{eigen}. It is $\Delta n_{hp}(k)=-\left( 
\frac{\omega _0}{\omega _0^2+k^2}+\frac{2\omega _0}{\omega _0^2+4k^2}+\frac{%
6\omega _0}{9\omega _0^2+4k^2}\right) \frac 2\pi $. Using this, we get

\begin{equation}
\label{A3}I_{hp}=\frac{15}2\omega _0^3 
\end{equation}

\subsubsection{End Crossing}

The product that we wish to evaluate is

\begin{equation}
\label{A4}I_{end}=\left( \frac{\prod\limits_k\varepsilon _k}{%
\prod\limits_{k>2}\varepsilon _k^{\ddagger }}\right) ^{1/2} 
\end{equation}
This infinite product in the above equation is over the continuous spectra
of the two Hamiltonians and may be evaluated. The change in the density of
states is now just half of the density of states density of states for the
hairpin case. That is, $\Delta n_{end}(k)=\frac 12\Delta n_{hp}(k)$. At the
first sight, this leads to a problem, as $\int_0^\infty dk$ $\Delta
n_{end}(k)=-3/2$, instead of the expected $2$ (as $\widehat{H}^{\ddagger }$
has two bound states while $\widehat{H}^{ms}$ has none). The solution to ths
is quite well known - $\widehat{H}^{ms}$ has a state with eigenvalue $\omega
_0^2$ where its continuous spectrum starts, and half of this state is to be
considered as a bound state. Then, we can write the above as:

$$
I_{end}=\exp \left( -\frac 12\int_0^\infty dk\Delta n_{end}(k)\ln \left(
\omega _0^2+k^2\right) \right) \sqrt{\omega _0} 
$$

Hence we find

\begin{equation}
\label{A6}I_{end}=\left( \frac{15}2\right) ^{1/2}\omega _0^2 
\end{equation}

\large
\pagebreak
. \vspace{1in} 
\begin{figure}
%\special{eps:Fig1.eps x=1in y=1in} 
\caption {  The potential energy per segment of the chain, plotted as a function of position} \label{fig1}
\end{figure}

%\pagebreak
\vspace{1in} 
\begin{figure}
%\special{eps:Fig2.eps x=1in y=1in} 
\caption{The potential energy along the reaction co-ordinate.  The $E_{act}$ is independent of
the length of the chain.  After the barrier is crossed, there is a region of
width $W$, with $W$ proportional to $N$, which is to be crossed.  The time required to cross this regions is 
$t_{cross}$} \label{fig2}
\end{figure}

%\pagebreak
\vspace{1in} 
\begin{figure}
%\special{eps:Fig3.eps x=1in y=1in} 
\caption{The transition state} \label{fig3}
\end{figure}

%\pagebreak
\vspace{1in} 
\begin{figure}
%\special{eps:Fig4.eps x=1in y=1in} 
\caption{End Crossing} \label{fig3a}
\end{figure}

%\pagebreak
\vspace{1in} 
\begin{figure}
 %\special{eps:Fig5.eps x=1in y=1in} 
\caption{Hairpin Crossing} \label{fig3b}
\end{figure}

%\pagebreak
\vspace{1in} 
\begin{figure}
%\special{eps:Fig6.eps x=1in y=1in} 
\caption{The barrier and its inverted form.  The barrier heights in the
 forward and backward directions are shown.  The dotted line represents
 the path that determines the activation energy} \label{fig4}
\end{figure}

%\pagebreak
\vspace{1in} 
\begin{figure}
 %\special{eps:Fig7.eps x=1in y=1in}
\caption{The free energy per segment of the polymer, shown as a function of the
 position of the segment (for the case where DNA is drawn through a pore).
 As the segment goes from the left (-ve) to right (+ve), the free
energy changes by -$\Delta V$} \label{fig5}
\end{figure}

%\pagebreak
\vspace{1in} 
\begin{figure}
%\special{eps:Fig8.eps x=1in y=1in}
\caption{The full curve is the plot of the potential for the motion of the particle 
for $0<n<s$, while the dotted curve is the potential for $s<n<N$. 
The particle starts at $R_0$ at the time $t=0$, moves on the full curve and reaches 
$R_s$ at the time $n=s$. At this time, the potential suddenly switches to its negative. 
The particle then moves on this potential (dotted curve). The path of the particle is 
drawn with dashes and direction in which it moves is shown by the arrows.}
\label{fig6}
\end{figure}

%\pagebreak
\vspace{1in} 
\begin{figure}
%\special{eps:Fig9.eps x=1in y=1in} 
\label{fig7}
\caption{The transition state for a hydrophilic chain with a 
hydrophobic signal sequence, passing through a hydrophobic pore.
Compare with figure 14-14 of the book by Alberts et. al.
}
\end{figure}

%\pagebreak
\vspace{1in} 
\begin{figure}
 %\special{eps:Fig10.eps x=1in y=1in} 
\caption{This is not a possible transition state.} \label{fig8}
\end{figure}


\begin{references}
\bibitem{text1}  This is a consequence of assuming a completely flexible
model for the polymer. This result is not true for a semi-flexible polymer.
\end{references}
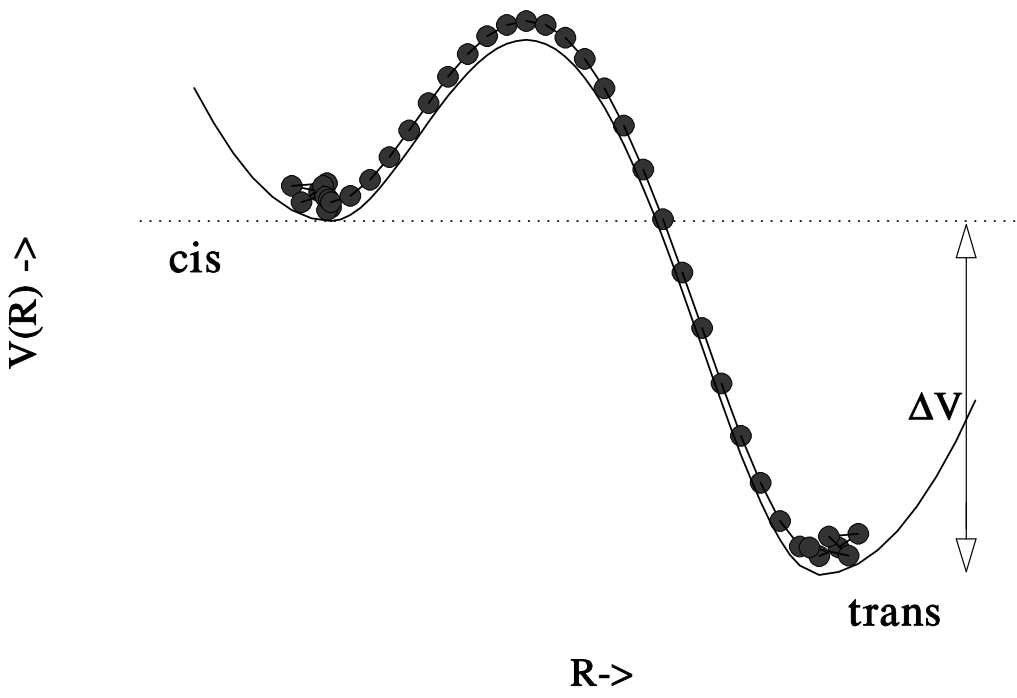
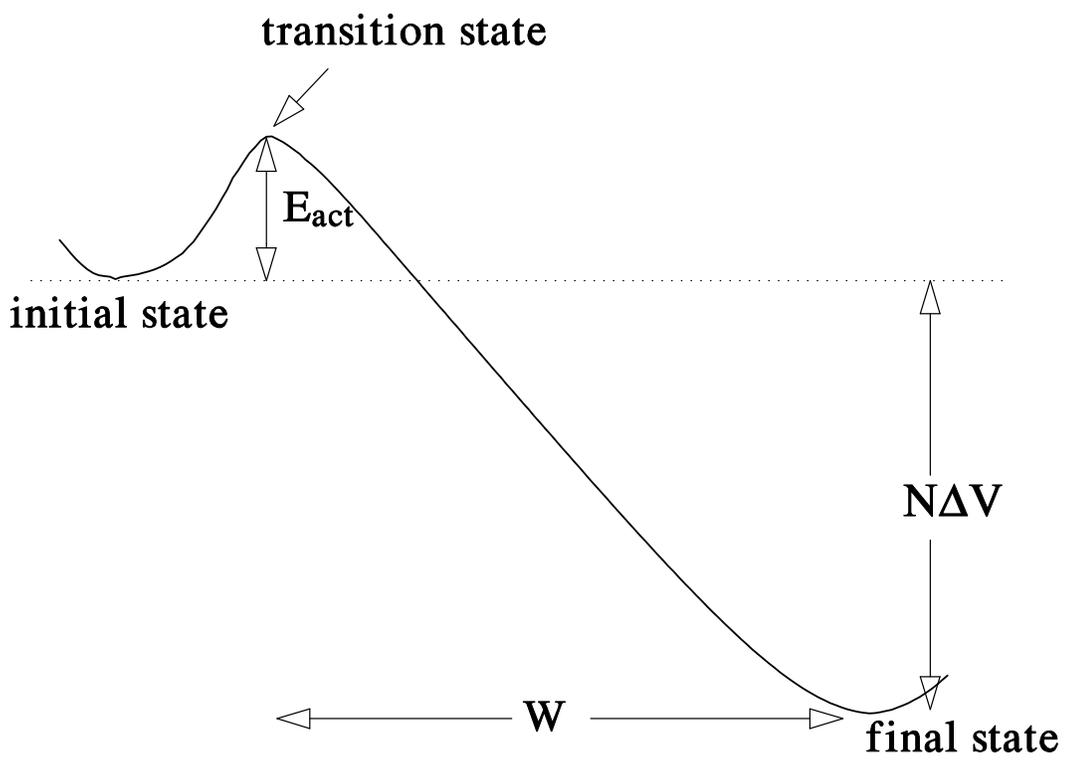
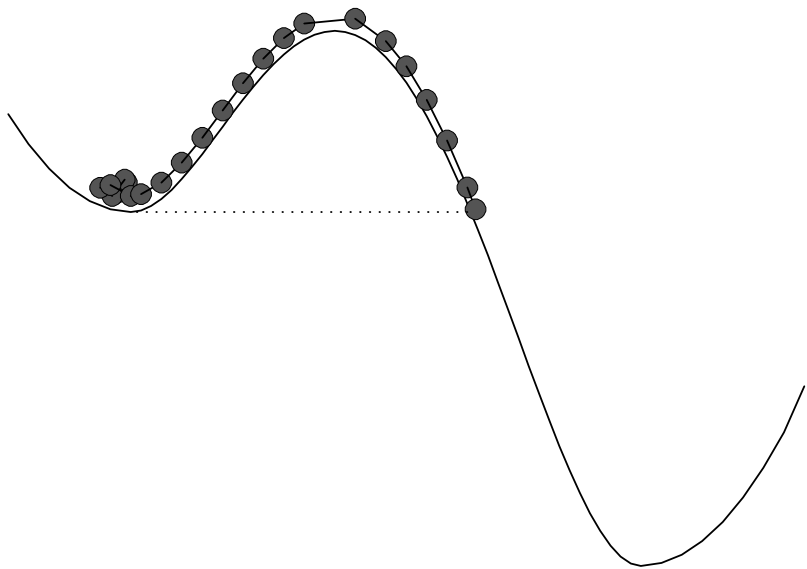
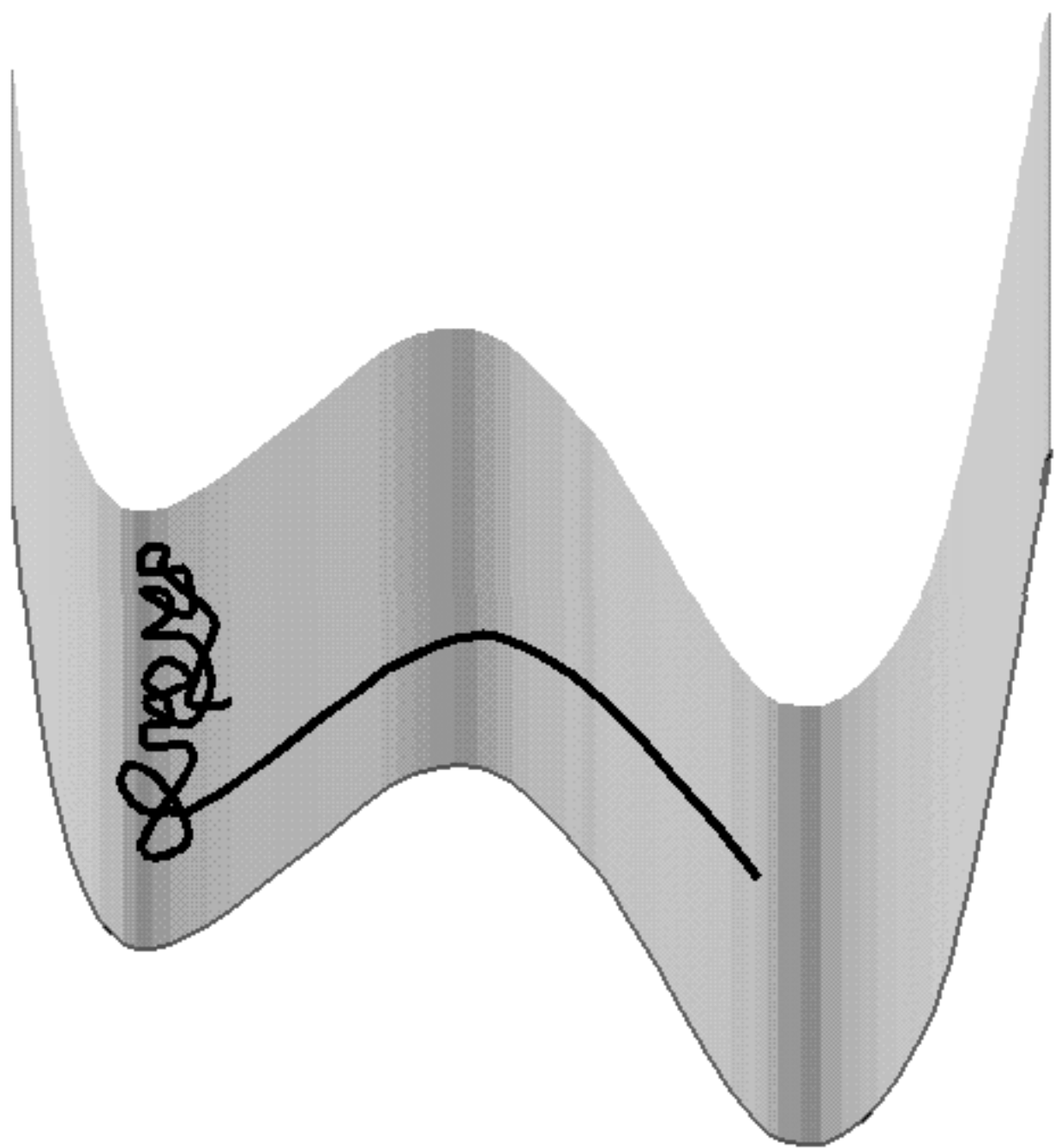
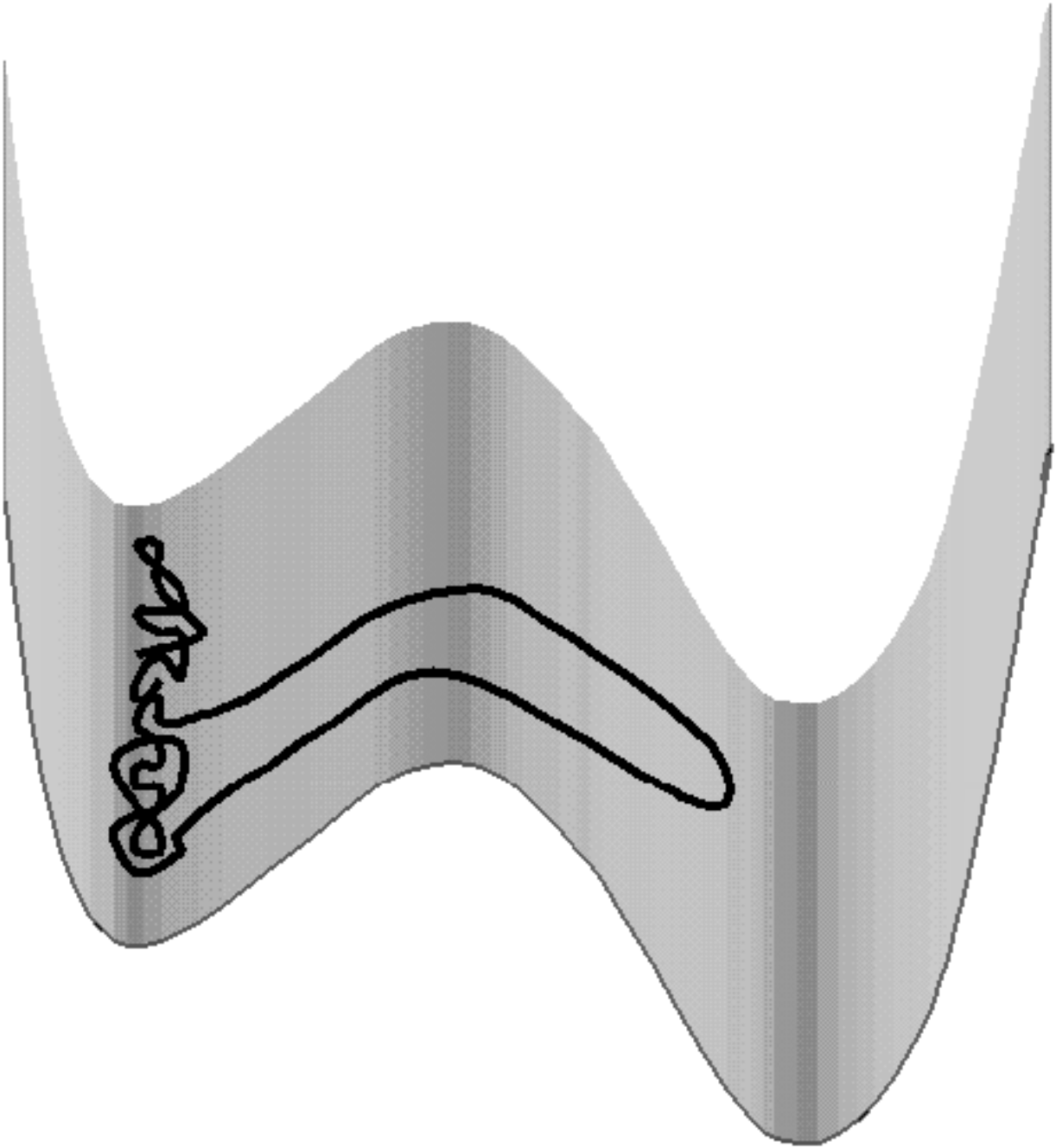
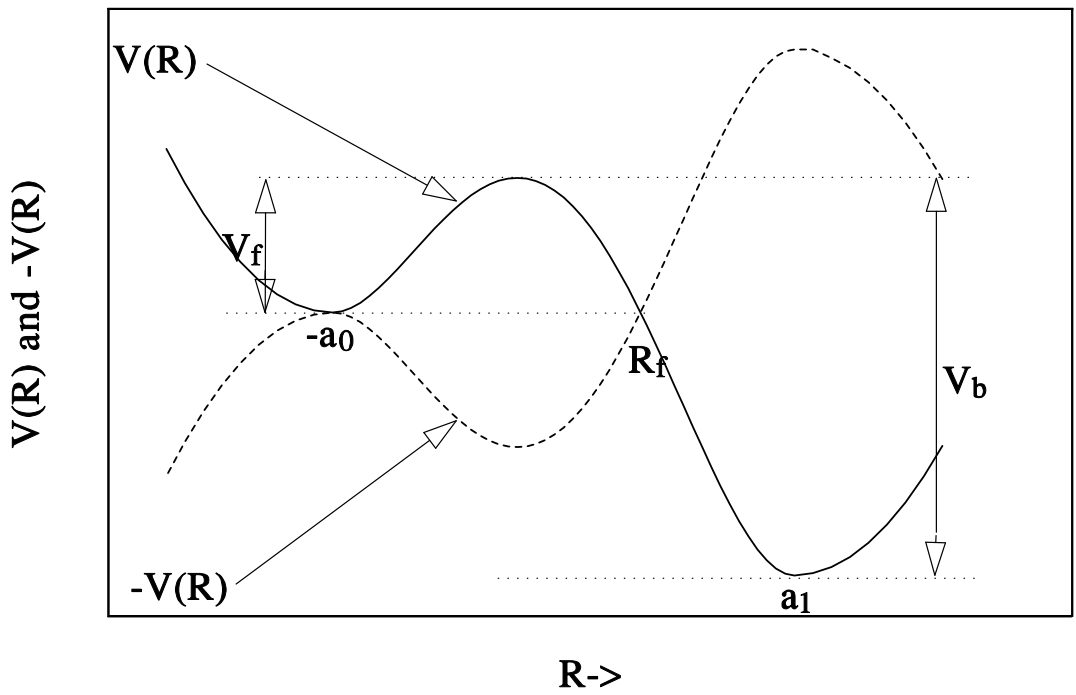
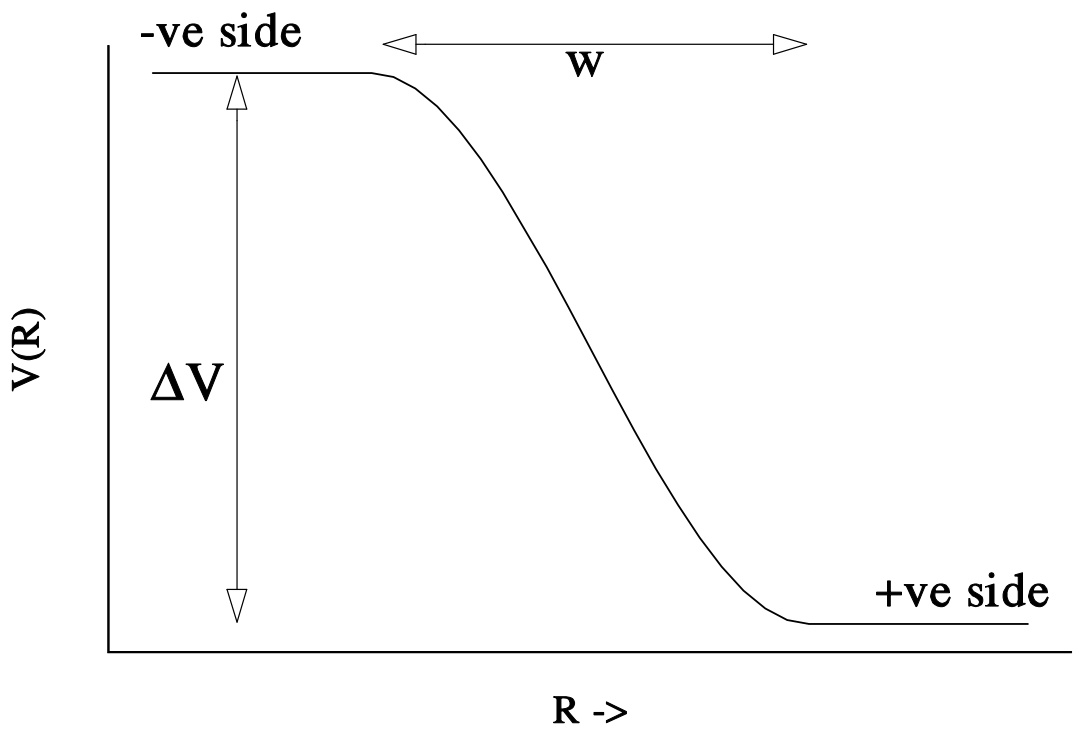
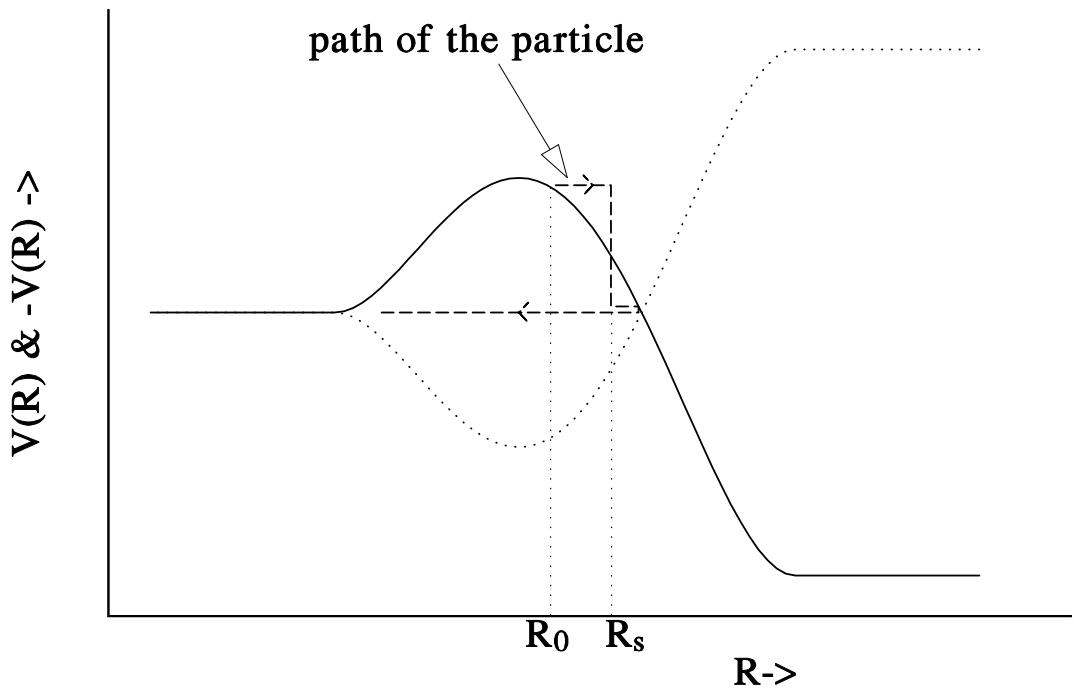
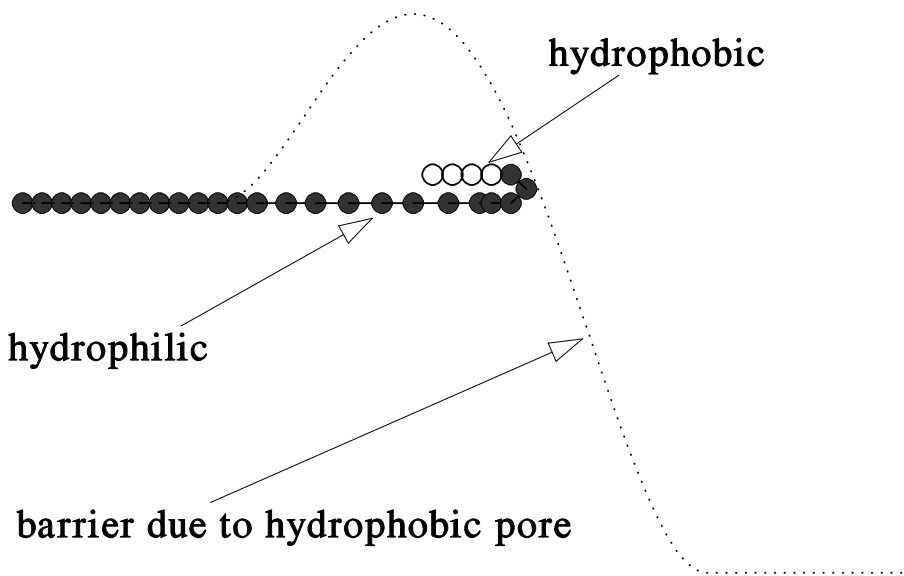
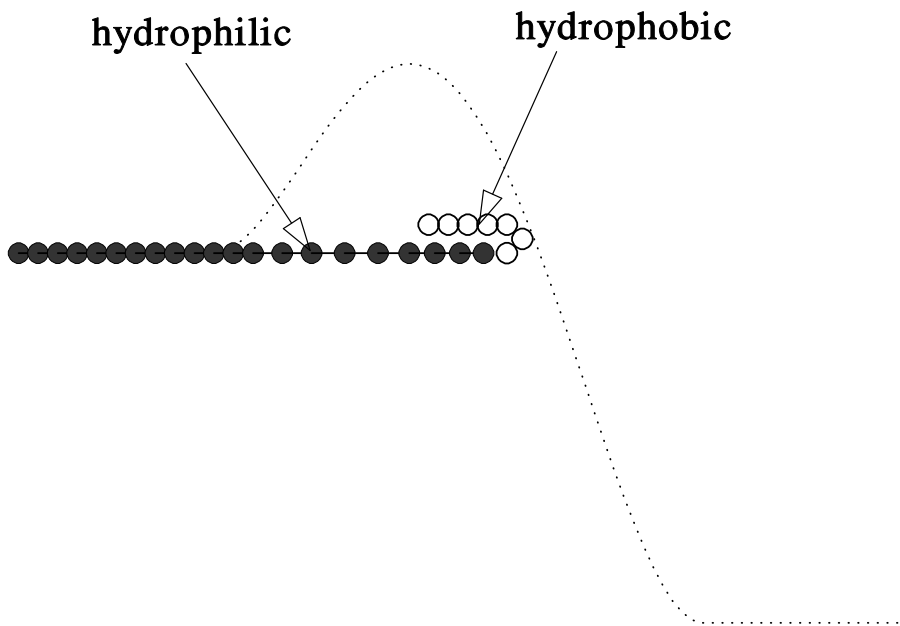
\end{document}